\documentclass[numberedappendix, apj, iop]{emulateapj}

\slugcomment{Submitted to ApJ, v3}
\shorttitle{The NS Mass Distribution}
\shortauthors{K{\i}z{\i}ltan et al.}
\submitted{Submitted to the Astrophysical Journal}
\usepackage[hyperindex , citecolor=blue, colorlinks=true, linkcolor=linkcolor, urlcolor=true, bookmarks=true]{hyperref}
\usepackage{psfrag}
\usepackage{auto-pst-pdf}
\hypersetup{colorlinks=true, linkcolor= magenta}
\usepackage{amssymb, amsmath, subfigure}
\def\be{\begin{equation}}
\def\ee{\end{equation}}
\def\bee{\begin{eqnarray}}
\def\eee{\end{eqnarray}}
\def \msun {\,\text{M}_{\odot} } 		

\def\Sref#1{\S\ref{Sec:#1}}
\def\Fref#1{Figure~\ref{Fig:#1}}
\def\Tref#1{Table~\ref{Tab:#1}}

\newcommand{\ppd}{$\text{P}$-$\dot{\text{P}}$ }

\psfrag{xlab}[c][c][1.1]{Neutron star mass [$\mathrm{M}_\odot$]}
\psfrag{ylab}[c][c][1.1]{Density}

\psfrag{xlab2}[c][c][1.1]{Neutron star mass [$\mathrm{M}_\odot$]}
\psfrag{ylab2}[c][c][1.1]{Density}

\begin{document}

\shorttitle{The Neutron Star Mass Distribution}
\shortauthors{K{\i}z{\i}ltan et al.}
\title{The Neutron Star Mass Distribution}
\author{B\"ulent K{\i}z{\i}ltan\altaffilmark{1,2}, Athanasios Kottas\altaffilmark{3}, Maria De Yoreo\altaffilmark{3} \& Stephen E. Thorsett\altaffilmark{2, 4}}
\altaffiltext{1}{Harvard-Smithsonian Center for Astrophysics, 60 Garden Street, Cambridge, MA 02138; e-mail: bkiziltan@cfa.harvard.edu}
\altaffiltext{2}{Department of Astronomy \& Astrophysics, University of California \& UCO/Lick Observatory, Santa Cruz, CA 95064}
\altaffiltext{3}{Department of Applied Mathematics and Statistics, University of California, Santa Cruz, CA 95064}
\altaffiltext{4}{Department of Physics, Willamette University, Salem, OR 97031}

\begin{abstract}
In recent years, the number of pulsars with secure mass measurements has increased to a level that allows us to probe the underlying neutron star (NS) mass distribution in detail. We critically review the radio pulsar mass measurements. For the first time, we are able to analyze a sizable population of NSs with a flexible modeling approach that can effectively accommodate a skewed underlying distribution and asymmetric measurement errors. We find that NSs that have evolved through different evolutionary paths reflect distinctive signatures through dissimilar distribution peak and mass cutoff values. NSs in double neutron star and neutron star-white dwarf systems show consistent respective peaks at $1.33\msun$ and $1.55\msun$ suggesting significant mass accretion ($\Delta m\approx 0.22\msun $) has occurred during the spin-up phase. The width of the mass distribution implied by double NS systems is indicative of a tight initial mass function while the inferred mass range is significantly wider for NSs that have gone through recycling. We find a mass cutoff at $\sim2.1\msun$ for NSs with white dwarf companions which establishes a firm lower bound for the maximum NS mass. This rules out the majority of strange quark and soft equation of state models as viable configurations for NS matter. The lack of truncation close to the maximum mass cutoff along with the skewed nature of the inferred mass distribution both enforce the suggestion that the 2.1$\msun$ limit is set by evolutionary constraints rather than nuclear physics or general relativity, and the existence of rare super-massive NSs is possible. 

\end{abstract}

\section{Introduction}\label{Sec:intr}

The mass of a neutron star (NS) has been a prime focus of compact objects astrophysics since the discovery of neutrons. Soon after Chadwick's {\em Letter} on the ``Possible existence of a neutron'' (\citeyear{Chadwick:32}), heated discussions around the world started to take place on the potential implications of the discovery. In 1932, during one of these discussions in Copenhagen, Landau shared his views with Rosenfeld and Bohr where he anticipated the existence of a dense-compact star composed primarily of neutrons \citep[e.g., ][p. 242]{Shapiro:83}. The prediction was not officially announced until \citeauthor{Baade:34b} published their work where the phrase ``neutron star'' appeared in the literature for the first time \citep{Baade:34b}. Their following work explained the possible evolutionary process leading to the production of a NS and the physics that simultaneously constrains the mass and radius in more detail \citep{Baade:34a, Baade:34}. 

The ensuing discussions were primarily focused on the mass of these dense objects. In 1931, \citeauthor{Chandrasekhar:31} had already published his original work in which he calculates the upper mass limit of an ``ideal'' white dwarf as $0.91\msun$, while the following year, Landau intuitively predicted that a limiting mass should exist close to $1.5\msun$ \citep{Landau:32}. Following the works of \citeauthor{Chandrasekhar:31} and \citeauthor{Landau:32}, and using the formalism developed by \citeauthor{Tolman:39}, \citeauthor{Oppenheimer:39} predicted an upper mass limit for NSs to be $0.7$--$3.4\msun$ \citep{Tolman:39, Oppenheimer:39}.

Since then, continuing discussions on the mass range a NS can attain have spawned a vast literature \citep[e.g.,][and references therein]{Rhoades:74, Joss:76, Thorsett:99, Baumgarte:00, Schwab:10}.
 
Masses of NSs at birth are tuned by the intricate details of the astrophysical processes that drive core collapse and supernova explosions \citep{Timmes:96}. The birth mass is therefore of particular interest to those who study these nuclear processes. An earlier attempt by \cite{Finn:94} finds that NSs should predominantly fall in the $1.3$--$1.6\msun$ mass range. The most comprehensive work to date by \cite{Thorsett:99} finds that the mass distribution of observed pulsars are consistent with $M=1.38^{-0.06}_{+0.10}\msun$, a remarkably narrow mass range. The recent work of \cite{Schwab:10} on the other hand, argues that there is evidence for multi-modality in the NS birth mass distribution (for discussion, see \Sref{disc}). 

 The maximum possible mass of a NS has attracted particular attention because it delineates the low mass limit of stellar mass black holes \citep{Rhoades:74, Fryer:01}. When combined with measurements of NS radii, it also provides a distinctive insight into the structure of matter at supranuclear densities \citep{Cook:94, Haensel:03, Lattimer:04, Lattimer:07}. Although more modern values theoretically predict a maximum NS mass of $M_{max}\approx2.2$--$2.9\msun$ \citep{Bombaci:96, Kalogera:96, Heiselberg:00}, it is still unclear whether very stiff equations of states (EOSs) that stably sustain cores up to the general relativity limit ($\sim3\msun $) can exist. 


Recent observations of pulsars in the Galactic plane as well as globular clusters suggest that there may be, in fact, NSs with masses significantly higher than the canonical value of $1.4\msun$ \citep[e.g.,][]{Champion:08, Ransom:05, Freire:08, Freire:08b, Freire:08a}. NSs in X-ray binaries also show systematic deviations from the canonical mass limit \citep[e.g.,][]{van-Kerkwijk:95, Barziv:01, Quaintrell:03, van-der-Meer:05, Ozel:09, Guver:10}.

The most precise measurements of NS masses are achieved by estimating the relativistic effects to orbital motion in binary systems. The exquisite precision of these mass measurements presents also a unique means to test general relativity in the ``strong field'' regime \citep[e.g.,][]{Damour:92, Psaltis:08}. As the masses of NSs also retain information about the past value of the effective gravitational constant $G$, with the determination of the NS mass range it may be even possible to probe the potential evolution of such physical constants \citep{Thorsett:96}.

A comprehensive insight into the underlying mass distribution of NSs thus provides not only the means to study NS specific problems. It also offers diverse sets of constraints that can be as broad as the high-mass star formation history of the Galaxy \citep{Gould:00}, or as particular as the compression modulus of symmetric nuclear matter \citep{Glendenning:86, Lattimer:90}. 

The present work aims to set up a framework by which we can probe the underlying mass distributions implied by radio pulsar observations. We develop a Bayesian framework that not only allows more flexibility for the inferred distribution but also accommodates asymmetric measurement errors in full parametric form. Unlike conventional statistical methods, with a Bayesian approach it is possible to separately infer peaks, shapes and cutoff values of the distribution with appropriate uncertainty quantification. This gives us unique leverage to probe these parameters which separately trace independent astrophysical and evolutionary processes.

In order to prevent contamination of the population, which may lead to systematic deviations from the probed mass distribution, we keep the observed pulsar sample as uniform as possible. We choose mass measurements that do not have strong {\em a priori} model dependencies and therefore can be considered secure.

In \Sref{theo} we review theoretical constraints on NS masses. We derive useful quantities such as the NS birth mass $M_{birth}$, the amount of mass expected to be transferred onto the NS primary during recycling $\Delta m_{acc}$, and the viable range of maximum mass (upper 97.5\% probability) cutoff value $M_{max}$ for NSs. The observations are reviewed in \Sref{obs}. We describe the statistical approach used to probe the underlying NS mass distribution in \Sref{esti}. After we summarize in \Sref{summ}, the range of implications and following 
conclusions are discussed in \Sref{disc}. For brevity, the details of the statistical 
model, the method for inference, and results from model checking are included 
in the Appendix \Sref{inf} and \Sref{postpred}.

\section{Theoretical Constraints}\label{Sec:theo}
This section summarizes theoretical estimates on the birth mass of NSs, the amount of mass that can be accreted onto NSs, and the constraints on the maximum NS mass.

	\subsection{Birth Mass}\label{Sec:birth}

The canonical mass limit $M_{ch}\sim1.4 \msun $ is the critical mass beyond which the degenerate remnant core of a massive star or a white dwarf will lose gravitational stability and collapse into a NS. This limiting mass is an approximation which is sensitive to several nuclear, relativistic and geometric effects \citep[see][for review]{Ghosh:07, Haensel:07}. In addition to these effects, the variety of evolutionary processes that produce NSs warrant a careful treatment.

A more precise parametrization of the Chandrasekhar mass is
\be
M_{ch}=5.83\, Y_{e}^{2}\msun
\ee
where $Y_{e}=n_{p}/(n_{p}+n_{n})$ is the electron fraction. A perfect neutron-proton equality ($n_{p}=n_{n}$) with $Y_{e}=0.50$ yields a critical mass of
\be
M_{ch}=1.457\msun.
\ee
However, we have a sufficiently good insight into the processes that affect $M_{ch}$. So, we can go beyond the idealized cases and estimate the remnant's expected initial mass more realistically. 

The inclusion of more reasonable electron fractions ($Y_{e}<0.50$) yields smaller values for $M_{ch}$. General relativistic implications, surface boundary pressure corrections, and the reduction of pressure due to non-ideal Coulomb interactions ($e^{-}$-$e^{-}$ repulsion, ion-ion repulsion and $e^{-}$-ion attraction) at high densities all reduce the upper limit of $M_{ch}$. 

On the other hand, the electrons of the progenitor (i.e., white dwarf or the core of a massive star) material are not completely relativistic. This reduces the pressure leading to an increase in the amount of mass required to reach the gravitational potential to collapse the star. Finite entropy corrections and the effects of rotation will also enhance the stability for additional mass. These corrections, as a result, yield a higher upper limit for $M_{ch}$.

The level of impact on the birth masses due to some of these competing effects is not well constrained as the details of the processes are not well understood. An inclusion of the effects that are due to the diversity in the evolutionary processes alone requires a $\approx 20$\% correction \citep[for a detailed numerical treatment see][]{Butterworth:75} and therefore implies a broader mass range, i.e. $M_{ch}\sim1.17$--$1.75\msun$. 

The measured masses, however, are the effective gravitational masses rather than a measure of the baryonic mass content. After applying the quadratic correction term $M_{baryon}-M_{grav}\approx 0.075\, M_{grav}^{2}$ \citep{Timmes:96}, one can get 
\be
M_{birth}\sim1.08\text{--}1.57\msun 
\label{eq:birth}
\ee
as a viable range for gravitational NS masses which is believed to encapsulate the range of NS birth masses.

\subsection{Accreted Mass}\label{Sec:accr}

There is considerable evidence that at least some millisecond pulsars have evolved from a first generation of NSs which have accumulated mass and angular momentum from their evolved companion \citep{Alpar:82, Radhakrishnan:82, Wijnands:98, Markwardt:02, Galloway:02, Galloway:05}. There is also a line of arguments that support the possibility of alternative evolutionary processes that may enrich the millisecond pulsar population \citep{Bailyn:90, Kiziltan:09}. 

Possible production channels for isolated millisecond pulsars are mergers of compact primaries or accretion induced collapse (AIC). In the case where a NS is produced via AIC, the final mass configuration of the remnant is determined by the central density of the progenitor (C-O or O-Ne white dwarf) and the speed at which the conductive deflagration propagates \citep{Woosley:92}. 

While there are uncertainties for the parameters that describe the ignition and flame propagation, a careful treatment of the physics that tune the transition of an accreting white dwarf yields a unique baryonic mass $M_{baryon}\approx 1.39\msun$ for the remnant which gives a gravitational mass of $M_{grav}\sim 1.27\msun$ for NSs produced via AIC \citep{Timmes:96}. There is indirect evidence that the occurrence rate of AICs can be significant \citep{Bailyn:90}.

The physics of these production channels are still not understood well enough to make quantitative predictions of the NS mass distribution produced via these processes. But we can estimate the mass required to spin NSs up to millisecond periods by using timescale and angular momentum arguments.

For low mass X-ray binaries (LMXBs) accreting at typical rates of $\dot{m}\sim10^{-3}\dot{\text{M}}_{\text{Edd}}$, the amount of mass accreted onto a NS in $10^{10}$yr is $\Delta m \approx0.10\msun$. One can also estimate the amount of angular momentum required to spin the accreting progenitor up to velocities that equal the Keplerian velocity at the co-rotation radius. In order to transfer sufficient angular momentum ($L=I\times\omega$) and spin up a normal pulsar ($R\approx$12 km, $I\approx1.4\times10^{45}$ g\,cm$^{2}$) to millisecond periods, an additional mass of $\Delta m \approx0.20\msun $ is required. Hence,
\bee
\Delta m_{acc} \approx 0.10\text{--}0.20\msun 
\eee 
will be sufficient to recycle NS primaries into millisecond pulsars.

\subsection{Maximum Mass}\label{Sec:maxi}

The mass and the composition of NSs are intricately related. One of the most important empirical clues that would lead to constraints on a wide range of physical processes is the maximum mass of NSs. For instance, secure constraints on the maximum mass provide insight into the range of viable EOSs for matter at supranuclear densities. 

A first order theoretical upper limit can be obtained by numerically integrating the Oppenheimer-Volkoff equations for a low-density EOS at the lowest energy state of the nuclei \citep{Baym:71}. This yields an extreme upper bound to the maximum mass of a NS at $M_{max}\sim 3.2\msun$ \citep{Rhoades:74}. Any compact star to stably support masses beyond this limit requires stronger short-range repulsive nuclear forces that stiffens the EOSs beyond the causal limit. For cases in which causality is not a requisite ($v \rightarrow \infty$) an upper limit still exist in general relativity $\approx 5.2\msun$ that considers uniform density spheres \citep{Shapiro:83}. However, for these cases the extremely stiff EOSs that require the sound speed to be super-luminal ($dP/d\rho\geq c^{2}$) are considered non-physical. 

Differentially rotating NSs that can support significantly more mass than uniform rotators can be temporarily produced by binary mergers \citep{Baumgarte:00}. While differential rotation provides excess radial stability against collapse, even for modest magnetic fields, magnetic braking and viscous forces will inevitably bring differentially rotating objects into uniform rotation \citep{Shapiro:00}. Therefore, radio pulsars can be treated as uniform rotators when calculating the maximum NS mass.

While general relativity along with the causal limit put a strict upper limit on the maximum NS mass at $\sim3.2\msun$, the lower bound is mostly determined by the still unknown EOS of matter at these densities and therefore is not well constrained. There are modern EOSs with detailed inclusions of nuclear processes such as kaon condensation and nucleon-nucleon scattering which affect the stiffness. These EOSs give a range of $1.5$--$2.2\msun$ as the lower bound for the maximum NS mass \citep{Thorsson:94, Kalogera:96}. Although these lower bounds for a maximum NS mass are implied for a variation of more realistic EOSs, it is still unclear whether any of these values are favored. Therefore, 
\bee
M_{max} \sim 1.5\text{--}3.2\msun 
\eee
can be considered a secure range for the maximum NS mass value.

\section{Observations}\label{Sec:obs}

The timing measurements of radio pulsations from NSs offer a precise means to constrain orbital parameters \citep{Manchester:77}. For systems where only five Keplerian orbital parameters (orbital period: $P_{b}$, projected semi-major axis: $x$, eccentricity: $e$, longitude and the time of periastron passage: $\omega_{0}$, $T_{0}$) are measured, individual masses of the primary ($m_{1}$) and secondary ($m_{2}$) stars, and the orbital inclination $i$ cannot be separately constrained. They remain instead related by the measured mass function $f$ which is given by
\bee
f=\frac{(m_{2}\, \text{sin}\,i)^{3}}{M^{2}}=\left(\frac{2\pi}{P_{b}}\right)^{2} x^{3}\text{T}_{\odot}^{-1}
\eee
where $M=m_{1}+m_{2}$ and masses are in solar units, the constant $\text{T}_{\odot}\equiv \text{G}\msun /\text{c}^{3}= 4.925490947\mu$s, and $x$ is measured in light seconds.

 \begin{deluxetable}{lllc} []
\tabletypesize{\footnotesize}
\small
\singlespace
\tablecolumns{4}
\tablewidth{0pt} 
\tablecaption{Double neutron star systems}
\startdata
\hline\hline
 & & &  \\ [-0.5ex]
Pulsar	&	Mass [$\text{M}_\odot$] 	& 	68\% central limits	& Refs.\tablenotemark{a}\\ [1ex]
\hline \\[-0.5ex]
	\multicolumn{4}{c}{Double neutron star binaries}\\[1ex]
\hline \\
J0737$-$3039 & & & [1] \\
pulsar A 	& 	1.3381 	&	 $\pm 0.0007$ 		& \\
pulsar B 	&	 1.2489	&	$\pm 0.0007$		&  \\
\quad total 	&	  2.58708     & 	$\pm 0.00016$			& \\ [1ex]
J1518+4904 &&& [2]\\
pulsar 	& 1.56 	& 	$+0.13/-0.44$ 		&\\
companion & 1.05	&	$+0.45/-0.11$		 & \\
\quad total 	& 2.61	 &	$\pm 0.070$		 & \\ [1ex]
B1534+12 & & & [3] \\
pulsar 		& 	1.3332 	& 	$\pm 0.0010$ 	  &\\
companion 	& 	1.3452 	& 	$\pm 0.0010$ 	  & \\
\quad total 		&	2.678428	 & 	$\pm 0.000018$		& \\ [1ex]
J1756$-$2251 &&& [4]\\
pulsar 		& 	1.40 		& 	$+0.02/-0.03$  	 & \\
companion 	& 	1.18 		& 	$+0.03/-0.02$ 	  & \\
\quad total 		& 	2.574		&	$\pm 0.003$	 & \\ [1ex]
J1811$-$1736 &&& [5, 6]\\
pulsar 		& 	1.56		& 	$+0.24/-0.45$		&  \\
companion 	& 	1.12	& 	$+0.47/-0.13$			&   \\
\quad total 	& 	2.57		 &	$\pm 0.10$ &  \\ [1ex]
J1829+2456 &&& [7]\\
pulsar 		& 	1.20	& 	$+0.12/-0.46$		&   \\
companion 	& 	1.40	& 	$+0.46/-0.12$		&  \\
\quad total 	& 	2.59		 & 	$\pm 0.02$		&   \\ [1ex]
J1906+0746 &&& [8, 9]  \\
pulsar 		& 	1.248 	& 	$\pm 0.018$	 	&   \\
companion 	& 	1.365		& 	$\pm 0.018$		&  \\
\quad total 		&	2.61 		& 	$\pm 0.02$			& \\ [1ex]
B1913+16 &&& [10, 11]\\
pulsar 		& 	1.4398	& $\pm 0.0002$ 	& \\
companion 	&	1.3886 	& $\pm 0.0002$ 	&  \\ 
\quad total 	 	& 	2.828378	&  $\pm 0.000007$ &\\ [1ex]
B2127+11C &&& [12, $\star$]\\
pulsar		&	 1.358 	&	 $\pm 0.010$ 	  &\\
companion 	&	 1.354 	&	 $\pm 0.010$	  & \\
\quad total	 	& 	2.71279	&	$\pm 0.00013$	 & \\ [1ex]
\hline
\enddata
\tablenotetext{a}{References:
1: \cite{Kramer:06},
2: \cite{Thorsett:99},
3: \cite{Stairs:02},
4: \cite{Faulkner:05},
5: \cite{Stairs:06},
6: \cite{Corongiu:07},
7: \cite{Champion:05},
8: \cite{Kasian:08},
9: \cite{Lorimer:06},
10: \cite{Weisberg:10},
11: \cite{Taylor:92},
12:  \cite{Jacoby:06},
$\star$: In globular cluster.
}
\label{Tab:doub}
\end{deluxetable}

For some binary systems, the timing residuals cannot be modeled with only Keplerian parameters when the effects of general relativity are measurable. In these cases, the gravitational influence can be parametrized as five potentially measurable post-Keplerian (PK) parameters which have similar interpretations \citep{Taylor:92}; (1) $\dot{\omega}$: advance of periastron (2) $\dot{P}_{b}$: orbital period decay (3) $\gamma$: time dilation-gravitational redshift (4) $r$: range of Shapiro delay (5) $s$: shape of Shapiro delay, where these are described by
\bee
\dot{\omega} & = & 3\left(\frac{P_{b}}{2\pi}\right)^{-5/3}\left(\text{T}_\odot M\right)^{2/3}\left(1-e^2\right)^{-1} ,
\eee
\bee
\dot{P}_b & = & -\frac{192\pi}{5}\left(\frac{P_{b}}{2\pi}\right)^{-5/3} \left(1+\frac{73}{24}e^2+\frac{37}{96}e^4\right)\times \nonumber\\
 & & (1-e^2)^{-7/2}\,\text{T}_\odot^{5/3}\,m_1m_2\,M^{-1/3} ,
 \eee
\bee
\gamma & = & e\left(\frac{P_{b}}{2\pi}\right)^{1/3}\text{T}^{2/3}_\odot M^{-4/3}m_2\left(m_1+2m_2\right) ,
\eee
\bee
r & = & \text{T}_\odot m_2 ,
\eee
\bee
s & = & x\left(\frac{P_{b}}{2\pi}\right)^{-2/3}\text{T}_\odot^{-1/3}M^{2/3}\,m_2^{-1}.
\label{Eq:PK}
\eee
A comprehensive review of the observational techniques and measurements can be found in \cite{Lorimer:04} and \cite{Stairs:06}.

\begin{deluxetable}{lllr}[]
\tabletypesize{\footnotesize}
\small
\singlespace
\tablecolumns{4}
\tablewidth{0pt} 
\tablecaption{Neutron star - white dwarf binary systems}
\startdata
\hline\hline
 & & &  \\ [-0.5ex]
Pulsar	&	Mass [$\text{M}_\odot$] 	& 	68\% central limits & Refs.\tablenotemark{a}	\\ [1ex]
\hline \\ [-0.5ex]
\multicolumn{4}{c}{Neutron star - white dwarf binaries}\\ [1.ex]
\hline \\
J0437$-$4715	&	1.76 		&	$\pm 0.20$				& [1] \\
J0621+1002	&	1.70		&	$+0.10/-0.17$				&[2]	\\	
J0751+1807	&	1.26		&	$\pm 0.14$			&[2]\\ 
J1012+5307 	& 	1.64 		& 	$\pm 0.22$ 	 	& [3]\\
J1141$-$6545	&	1.27		&	$\pm 0.01$				&[4]	\\
J1614$-$2230		&	1.97		&	$\pm 0.04$			&	 [5] \\ 	 
J1713+0747 	& 	1.53		& 	$+0.08/-0.06$ 	& [6]\\ 
J1802$-$2124	&	1.24 		 &	 $\pm 0.11$ 		&[7]\\ 
B1855+09 		& 	1.57 		& 	$+0.12/-0.11$ 		& [8] \\ 
J1909$-$3744	&	1.438	&	$\pm 0.024$			&  [9]	\\
B2303+46		&	1.38		&	$+0.06/-0.10$			&[10]\\ [1ex]
J0024$-$7204H 		&	1.48		&	$+0.03/-0.06$		& [$\dagger$, $\star$] 	\\
J0514$-$4002A 	&     1.49		& $+0.04/-0.27$  	& [$\dagger$, $\star$]	\\
B1516+02B 	& 2.10		&	$\pm 0.19$			&  [$\dagger$, $\star$]	\\
J1748$-$2446I	&	1.91		&	$+0.02/-0.10$	& [$\dagger$, $\star$]	 \\
J1748$-$2446J	&	1.79		&	$+0.02/-0.10$	& [$\dagger$, $\star$] \\
B1802$-$07  	&	1.26		 & 	$+0.08/-0.17$			& [10, $\star$] \\
B1911$-$5958A	&	1.40 		& 	$+0.16/-0.10$  		&[11, $\star$] \\ [1.ex]
\hline
\enddata
\tablenotetext{a}{References: 
1: \cite{Verbiest:08},
2: \cite{Nice:08},
3: \cite{Callanan:98},
4: \cite{Bhat:08},
5: \cite{Demorest:10}
6: \cite{Splaver:05},
7: \cite{Ferdman:10},
8: \cite{Nice:03},
9: \cite{Jacoby:05},
10:  \cite{Thorsett:99},
11: \cite{Bassa:06},
$\dagger$: P. Freire (personal communication), 
$\star$: In globular cluster.
}
\label{Tab:nswd}
\end{deluxetable}

In systems where at least two PK parameters can be measured, $m_{1}$ and $m_{2}$ may be individually determined. In rare cases, more than two PK parameters are measurable. These over-constrained systems present a unique means to test for consistent strong-field gravitational theories \citep{Taylor:89}. 

In \Tref{doub} and \Tref{nswd} we compile a comprehensive list of NS masses that are determined by the relativistic orbital phenomena reflected onto the binary system orbital parameters. We include the mass estimates along with the 68\% confidence limits which are plotted on \Fref{psr}. 

\begin{figure}[]
\includegraphics[width=0.49 \textwidth,angle=0, trim=2.cm .1cm .1cm .1cm, clip=true]{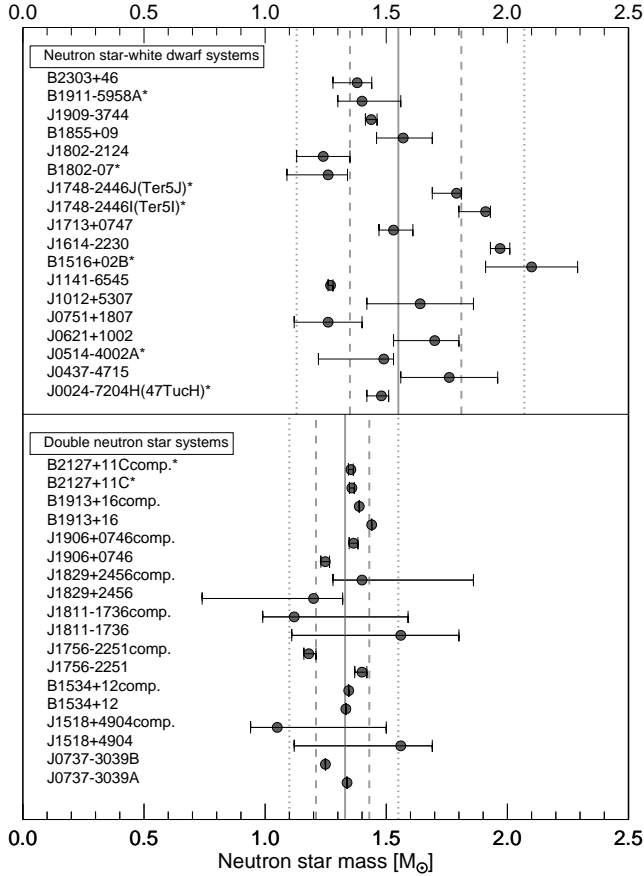} 	
\caption{Measured masses of radio pulsars. All error bars indicate the central 68\% confidence limits. Vertical solid lines are the peak values of the underlying mass distribution for DNS ($m=1.33\,\msun$) and NS-WD ($m=1.55\,\msun$) systems. The dashed and dotted vertical lines are the central 68\% and 95\% predictive probability intervals of the inferred mass distribution in \Fref{densities}. ``$\star$'' points to pulsars found in globular clusters.}
\label{Fig:psr}   
\end{figure} 

Pulsar surveys suffer from selection effects, especially in the low frequency ($<$1 GHz) regime. Recent surveys in 1.4 GHz have revealed a greater number of pulsars compared to previous surveys. The inverse square law, radio sky background, and propagation effects (i.e. pulse dispersion and scattering) in the interstellar medium also introduce global selection biases to the observed population. In addition to these biases affecting the population at large, there are also source dependent biases such as radio intermittency, pulse nulling, and the finite size of the emission beam. However, there is no evidence that these source dependent selection biases have mass dependent effects \citep[see][]{Lorimer:06a}. Therefore, we do not expect that these selection effects introduce mass dependent biases to the observed distribution.

We aim to prevent possible contamination of the sample with sub-populations which may have gone through different and not well understood evolutionary paths (e.g., isolated NSs). Even for the better constrained formation processes that lead to the production of DNS and NS-WD systems, theoretical models estimating the final NS masses are tentative.

\section{Estimating the Underlying Mass Distribution}\label{Sec:esti}

Recent advances in statistical methods have reached a level which allows us to extract information from sparse data with unprecedented detail. 

It can be clearly argued why modeling the underlying NS mass distribution as a single homogenous population is over-simplistic. There is no compelling line of reasoning that would require a single coherent (unimodal) mass distribution for NSs that we know have dissimilar evolutionary histories and possibly different production channels \citep[e.g., see][]{Podsiadlowski:04}. In fact, there is an increasing number of measurements that show clear signatures for masses that deviate from the canonical value of 1.4$\msun$. For instance, recent findings of \cite{van-Kerkwijk:10} imply that the mass for PSR B1957+20 may be as high as 2.4$\pm$0.12$\msun$. As we show in \Sref{model}, a flexible modeling approach can be used to test whether the implied masses belong to the same distribution. We argue that with the number of NS mass measurements available in \Tref{doub} and \Tref{nswd}, clear signatures should be manifest in the inferred underlying mass distributions if appropriate statistical techniques are utilized. Since we still operate in the sparse data regime, it is useful, if not necessary, to use Bayesian inference methods.

For the range of calculations we use mass measurements obtained directly from pulsar timing. The methods used for estimating NS masses other than radio timing, have intrinsically different systematics, and therefore require a more careful treatment when assessing the implied NS mass distribution. The inclusion of mass estimates of NSs in X-ray binaries along with these more secure measurements would potentially perturb the homogeneity of the sample and the coherence of the inference. 

For an all inclusive assessment of NS masses, more sophisticated hierarchical inference methods may be required. For sparse data, a proper statistical treatment of different systematic effects and {\em a priori} assumptions is not trivial. Also, the expected loss in precision may outweigh the gain obtained from a more detailed approach. Without properly tested and calibrated tools, further inclusion of NSs whose masses are not measured by pulsar timing in radio may just contaminate the sample and can therefore be misleading \citep[e.g., see][]{Steiner:10}.

\subsection{Statistical Model}
\label{Sec:model}

Here, we present the statistical model to estimate the NS mass 
distribution. The approach is based on a formulation that incorporates
errors in measurement of NS mass estimates. Specifically, the model formulation:
\begin{equation}\label{error-model}
m_{i} = \mathcal{M}_{i}+w_{i}, \,\,\,\,\, i=1,...,n
\end{equation}
where, for the $i$-th neutron star, $m_{i}$ is the estimate of the NS 
mass $\mathcal{M}_{i}$ and $w_{i}$ is the associated error. We thus 
need a model for the NS mass distribution and for the measurement 
error distribution.
Evidently, the key focus of inference is the NS mass distribution, but 
a flexible specification for the error distribution is needed to ensure
that this inference is not biased. 
At the same time, the model specification must take into account the 
limited amount of data. The proposed modeling approach achieves 
a balance between these considerations, and importantly, enables 
relatively straightforward implementation of inference through 
posterior simulation computational methods.

Visual inspection of the pulsar mass estimates (see \Tref{doub}, \Tref{nswd} 
and \Fref{psr}) suggests that skewness may be present in the NS mass 
distribution, at least for the NS-WD systems. It is therefore important 
to extend the normality assumption which is implicit in the existing estimation methods. Furthermore, it is clear from the 
error bars of the pulsar mass estimates that an asymmetric measurement
error distribution is needed for some of the observations, especially
for the DNS systems. The statistical model developed below allows for 
skewness both in the NS mass distribution and the error distribution 
while encompassing the normal distribution for either as a special case. 

The pulsars in less constrained systems (e.g., with only one post-Keplerian parameter 
determined) typically have asymmetric measurement errors. The flexibility of the 
statistical modeling approach developed here allows us to take full advantage of all 
available mass measurements in \Tref{doub} and \Tref{nswd}. It is noteworthy that the model is generic enough so it can be adopted to other similar astrophysical problems and serve as a useful reference.

Regarding the model for the NS mass distribution, we work with a
skewed normal distribution with density function given by 
\begin{equation}\label{skew-normal}
\text{SN}(\mathcal{M} \mid \mu, \sigma, \alpha)
= \frac{2}{\sigma} \phi \left( \frac{\mathcal{M} - \mu}{\sigma} \right)
\Phi \left( \frac{(\mathcal{M} - \mu) \alpha}{\sigma} \right)
\end{equation}
where $\phi(\cdot)$ and $\Phi(\cdot)$ denote the density function and cumulative
distribution function, respectively, of the standard normal distribution.
Here, $\mu \in \mathbb{R}$ is a location parameter, 
$\sigma \in \mathbb{R}^{+}$ is a scale parameter, and 
$\alpha \in \mathbb{R}$ is a skewness parameter. This model was studied by \cite{Azzalini:85} and is one of 
the more commonly used skewed normal distributions.
Note that $\alpha = 0$ yields the normal distribution 
with mean $\mu$ and standard deviation
$\sigma$ as a special case of (\ref{skew-normal}), which highlights the role
of $\alpha$ as a skewness parameter. In particular, positive/negative 
values of $\alpha$ result in right/left skewness for the density in 
(\ref{skew-normal}). Hence, an appealing feature of this model is that, 
within the context of Bayesian inference, we can make probabilistic 
assessments for skewness of the NS mass distribution relative to 
a normal distribution through, for instance, a posterior interval estimate 
for parameter $\alpha$. As discussed in \Sref{inference}, we find some evidence 
for skewness in the NS mass distribution corresponding to the NS-WD 
systems, but not for the DNS systems.

Next, we describe the model for the error distribution, which is 
motivated by the process used to produce the pulsar mass estimates
and the associated error bars. For each pulsar (either from a NS-WD or a DNS system) an empirical 
density curve for its mass is constructed based on how well the post-Keplerian 
parameters of the system can be constrained. We generically denote the 
final constructed density for the $i$-th pulsar
as $h_{i}(m)$, and note that, although it is unimodal, it may be asymmetric 
(especially for pulsars that are in a system for which only one post-Keplerian parameter can be constrained)
resulting in the asymmetric error bars reported for some of the systems 
in \Tref{doub} and \Tref{nswd}. The pulsar mass estimate, $m_{i}$, is obtained as
the mode of this density, whereas the error bars, $+u_{i}/-\ell_{i}$, 
define the interval, $(m_{i} - \ell_{i},m_{i} + u_{i})$, of the
smallest possible length with 68\% probability 
coverage. Numerically, the interval is obtained by starting from the
peak value of $h_{i}(m)$ and slicing down the density until 
$\int_{m_{i} - \ell_{i}}^{m_{i} + u_{i}} h_{i}(m) \, \text{d}m = 0.68$, which, 
given the unimodal shape of density $h_{i}(m)$, also implies 
that $h_{i}(m_{i} - \ell_{i}) = h_{i}(m_{i} + u_{i})$.

Hence, in the context of model (\ref{error-model}), the errors $w_{i}$ 
are realizations from a distribution with mode at 0. We thus seek 
a measurement error distribution which is parameterized in terms of 
its mode, allows asymmetry around the mode, and yields the normal 
distribution in the special case of symmetry. A particularly suitable
choice is the asymmetric normal distribution studied in \cite{Fernandez:98} with density function given by 
\begin{eqnarray}\label{skew-normal-errors}
\lefteqn{\text{AN}(w \mid c,d) =} \\
 & & \frac{2}{d(c+\frac{1}{c})} \left\{ \phi\left(\frac{w}{cd}\right)
1_{[0,\infty)}(w) + \phi\left(\frac{c w}{d}\right)1_{(-\infty,0)}(w) \right\}
\nonumber
\end{eqnarray}
where $c > 0$, $d > 0$, and $1_{A}(\cdot)$ denotes the indicator 
function of set $A$. The mode of this density is at 0, when $c=1$ it 
reduces to the normal density with mean 0 and standard deviation $d$, 
and when $c > 1$ ($c < 1$) it is right skewed (left skewed). Therefore,
$d$ plays the role of a scale parameter, whereas $c$ is the asymmetry 
parameter.

A practically important feature of the asymmetric normal density
in (\ref{skew-normal-errors}) is that it leads to straightforward 
estimation of parameters $c_{i}$ and $d_{i}$ for the $i$-th pulsar,
using the values of the 68\% central limits, $+u_{i}/-\ell_{i}$, 
in \Tref{doub} and \Tref{nswd}. First, from the condition
$\text{AN}(-\ell_{i} \mid c_{i},d_{i}) =$ $\text{AN}(u_{i} \mid c_{i},d_{i})$
we obtain 
$\phi(- c_{i} d_{i}^{-1} \ell_{i})=$ $\phi( c_{i}^{-1} d_{i}^{-1} u_{i})$, 
which in turn yields $c_{i}=$ $(u_{i}/\ell_{i})^{1/2}$. Next, with 
$c_{i}$ determined, we specify $d_{i}$ by solving numerically for
its value that satisfies 
$\int_{-\ell_{i}}^{u_{i}} \text{AN}(w \mid c_{i},d_{i}) \text{d}w = 0.68$.
Computing this equation involves normal distribution function
evaluations, and it can be easily shown that there is a unique
solution for $d_{i}$. 
Note that $c_{i}=1$ when $\ell_{i}=u_{i}$, and thus the error
distribution is normal for the pulsars with symmetric error bars.
However, for pulsars with asymmetric error bars, asymmetry is
introduced in the respective error distribution components.
The extent of the asymmetry depends on the relative magnitude of 
$\ell_{i}$ and $u_{i}$; for instance, the maximum value for the asymmetry parameter $c_{i}$ 
is $2.02$ corresponding to PSR J1518+4904's companion, and the 
minimum value is 
$0.38$ corresponding to PSR J0514--4002A.

\subsection{Inferring the Neutron Star Mass Distribution}
\label{Sec:inference}

For each of the NS-WD and DNS systems, the data vector
comprises $\text{data} =$ $\{ (m_{i},c_{i},d_{i}): i=1,...,n \}$,
with $(c_{i},d_{i})$ computed as detailed in Section \ref{Sec:model}.
Then, combining the models for the NS mass and error distributions 
in (\ref{skew-normal}) and (\ref{skew-normal-errors}), respectively, 
the hierarchical model for the data can be written as 
\begin{eqnarray}\label{hierarchical-model}
m_{i} \mid \mathcal{M}_{i} \, & \sim & \, \text{AN}(m_{i} - \mathcal{M}_{i} \mid c_{i},d_{i}), 
\,\,\, i=1,...,n 
\nonumber \\
\mathcal{M}_{i} \mid \mu,\sigma,\alpha \, & \sim & \,
\text{SN}(\mathcal{M}_{i} \mid \mu, \sigma, \alpha), \,\,\, i=1,...,n
\end{eqnarray}
That is, given the respective NS masses $\mathcal{M}_{i}$, the pulsar mass estimates $m_{i}$ arise  conditionally independently from the asymmetric normal response distribution with mode at $\mathcal{M}_{i}$. Here, the $\mathcal{M}_{i}$ are modeled as conditionally independent 
realizations, given parameters $(\mu,\sigma,\alpha)$, from the skewed 
normal NS mass distribution.

Now, the likelihood function for the NS mass distribution parameters 
arises from the hierarchical model in (\ref{hierarchical-model}) by 
marginalizing over the $\mathcal{M}_{i}$, that is, 
\begin{eqnarray} \label{likelihood}
\lefteqn{\mathcal{L}(\mu,\sigma,\alpha;\text{data})=} \nonumber\\
& & \prod_{i=1}^{n} \int \text{AN}(m_{i} - \mathcal{M}_{i} \mid c_{i},d_{i})
\text{SN}(\mathcal{M}_{i} \mid \mu, \sigma, \alpha) \, \text{d}\mathcal{M}_{i}.
\nonumber
\end{eqnarray}
The integral is readily available analytically only in the special case of a normal 
distribution for both the errors and NS masses ($c_{i}=1$ and 
$\alpha=0$, respectively). Therefore, in general, numerical maximization
of the likelihood function to obtain the maximum likelihood estimates 
for parameters $\mu$, $\sigma$ and $\alpha$ is not straightforward. 
Even more challenging is uncertainty quantification for the point estimates, 
and its subsequent effect on the NS mass density; this would require
large-sample asymptotic results the use of which is particularly
problematic to justify given the small number of observations from
both the DNS and NS-WD systems.

We thus employ the Bayesian approach to inference for the NS mass
distribution, using Markov chain Monte Carlo (MCMC) methods for
sampling from the posterior distribution of model parameters \citep{Gelman:03}.
Under the Bayesian approach, model (\ref{hierarchical-model}) is
completed with priors for the NS mass distribution parameters. 
Details on the prior distributions are given in Appendix \Sref{inf}.

In addition to providing a coherent probabilistic framework for inference, 
the Bayesian model formulation enables harnessing the full hierarchical 
structure in (\ref{hierarchical-model}). In particular, our posterior 
simulation method retains the individual NS masses $\mathcal{M}_{i}$ 
as part of the full parameter vector along with the NS mass distribution 
parameters $(\mu,\sigma,\alpha)$. In fact, we work with a
transformed version $(\mu,\tau^{2},\psi)$ of $(\mu,\sigma,\alpha)$, 
such that $\sigma^{2}=$ $\tau^{2}$ + $\psi^{2}$ and $\alpha=$ $\psi/\tau$,
with $\tau >0$ and $\psi \in \mathbb{R}$. This 
re-parameterization facilitates implementation of the computational 
method for posterior inference as an efficient Gibbs sampler algorithm.
Key in this direction is also a stochastic representation of the skewed 
normal distribution in (\ref{skew-normal}) as a mixture of normal 
distributions \citep{Henze:86}. The specific result is given in Appendix \Sref{inf}, which includes also the
technical details of the MCMC posterior simulation method.

\begin{figure}[t]
\centering
\includegraphics[width=3.1in,height=3.in, trim=.5cm .5cm .cm 1.5cm, clip=true]{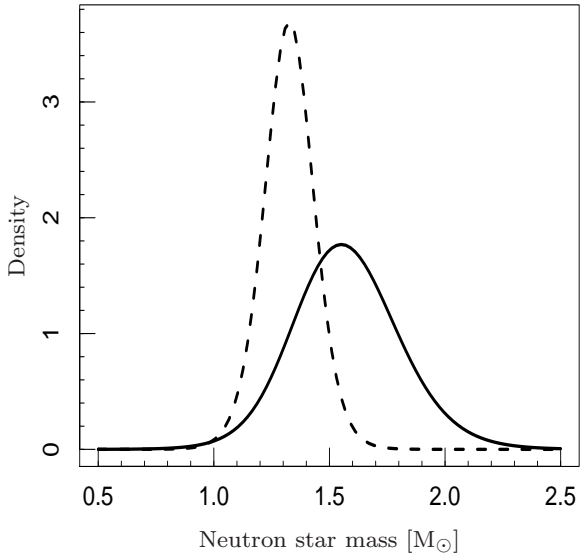} 	
\caption{Posterior predictive density estimates for the NS mass
distribution. DNS systems (dashed line) and NS-WD systems (solid 
line) mass densities have respective peaks at $1.33 \, \mathrm{M}_{\odot}$ 
and $1.55 \, \mathrm{M}_{\odot}$. The 68\% and 95\% posterior predictive 
intervals are given by $(1.21 \, \mathrm{M}_{\odot}, 1.43 \, \mathrm{M}_{\odot})$ 
and $(1.10 \, \mathrm{M}_{\odot}, 1.55 \, \mathrm{M}_{\odot})$ for the DNS systems,
and by $(1.35 \, \mathrm{M}_{\odot}, 1.81 \, \mathrm{M}_{\odot})$ and 
$(1.13 \, \mathrm{M}_{\odot}, 2.07 \, \mathrm{M}_{\odot})$ for the NS-WD systems.}
\label{Fig:densities}   
\end{figure} 

Implementing the Gibbs sampler described in Appendix \Sref{inf}, we 
obtain posterior samples for $(\mu,\tau^{2},\psi)$ and thus, through 
the transformation discussed above, for parameters $(\mu,\sigma,\alpha)$.
These samples can be used to explore a variety of inferences for the 
NS mass distribution. 

First, the point estimate for the density of the NS mass distribution 
is given by the posterior predictive density, 
$\mathcal{P}(\mathcal{M}_{0} \mid \text{data})$, where $\mathcal{M}_{0}$
denotes the (unknown) mass of a ``new'' unobserved pulsar which we 
seek to estimate (predict) given the observed data. The posterior 
predictive density is given by 
\begin{eqnarray}\label{pred-density}
\lefteqn{\mathcal{P}(\mathcal{M}_{0} \mid \text{data}) =} \nonumber \\ 
& & \int \text{SN}(\mathcal{M}_{0} \mid \mu, \sigma, \alpha) \, 
p(\mu,\sigma,\alpha \mid \text{data}) \, 
\text{d}\mu \, \text{d}\sigma \, \text{d}\alpha
\end{eqnarray}
where $p(\mu,\sigma,\alpha \mid \text{data})$ denotes the posterior 
distribution of the model parameters. Using expression (\ref{pred-density})
along with the samples from $p(\mu,\sigma,\alpha \mid \text{data})$, we can
compute through straightforward Monte Carlo integration the NS mass 
density estimates over any grid of mass values of interest. 
Figure~\ref{Fig:densities} plots the posterior predictive NS mass density 
estimates for the DNS and NS-WD systems, which have peaks at 
$1.33 \, \mathrm{M}_{\odot}$ and $1.55 \, \mathrm{M}_{\odot}$, respectively.
We can also sample from the posterior predictive distribution by 
sampling from the $\text{SN}(\mathcal{M}_{0} \mid \mu, \sigma, \alpha)$
distribution (using its normal mixture stochastic representation)
for each posterior sample of $(\mu,\sigma,\alpha)$.
The resulting samples quantify posterior predictive uncertainty around
the NS mass density peaks. In particular, for the DNS systems
the 68\% and 95\% posterior predictive intervals are 
$(1.21 \, \mathrm{M}_{\odot}, 1.43 \, \mathrm{M}_{\odot})$ 
and $(1.10 \, \mathrm{M}_{\odot}, 1.55 \, \mathrm{M}_{\odot})$,
whereas for the NS-WD systems the corresponding intervals are 
given by $(1.35 \, \mathrm{M}_{\odot}, 1.81 \, \mathrm{M}_{\odot})$ and 
$(1.13 \, \mathrm{M}_{\odot}, 2.07 \, \mathrm{M}_{\odot})$.

\begin{figure}[t]
\centering
\includegraphics[width=1.65in,height=1.65in, trim=.5cm .5cm .cm 1.5cm, clip=true]{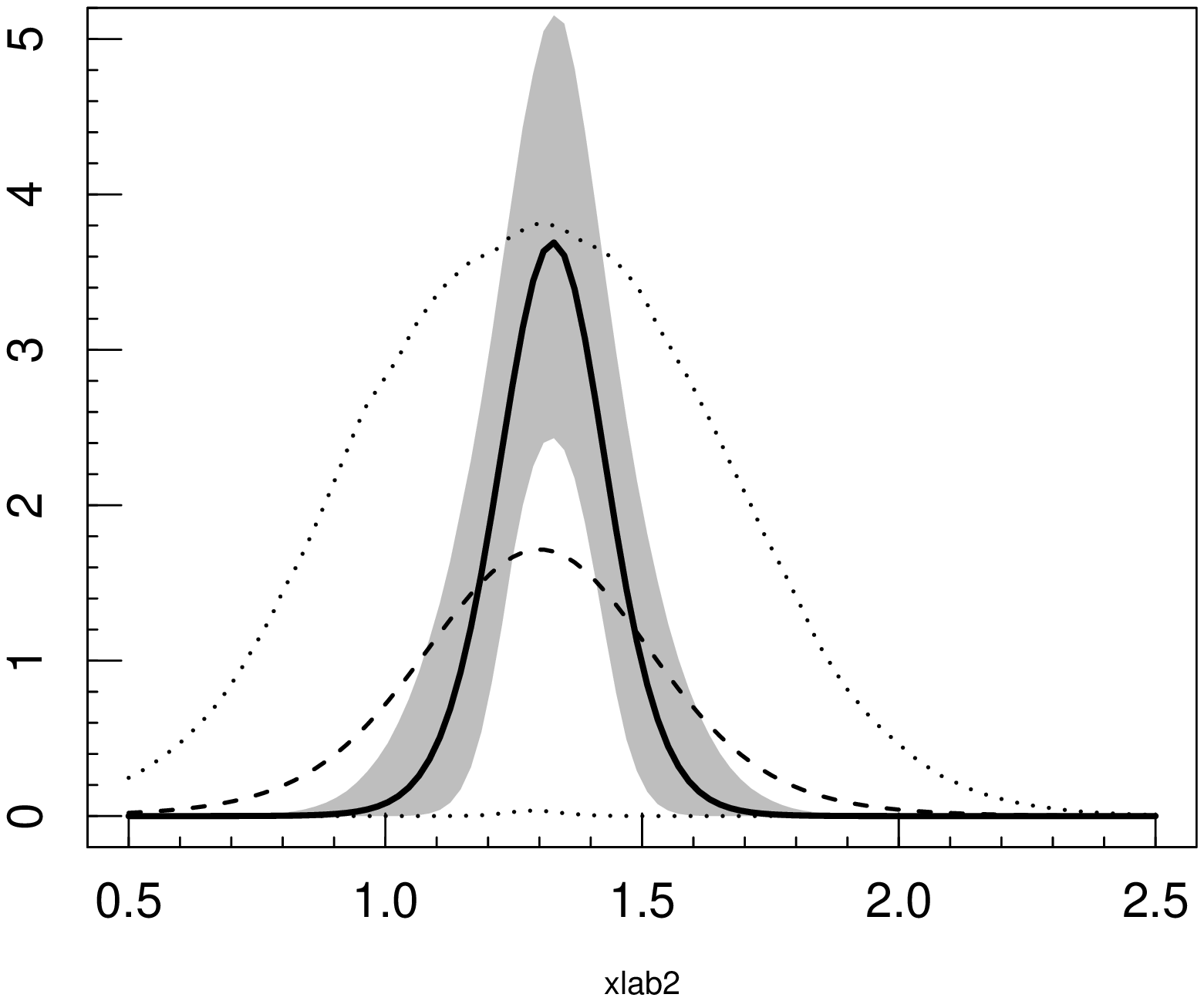}
\includegraphics[width=1.65in,height=1.65in,trim=.5cm .5cm .cm 1.5cm, clip=true]{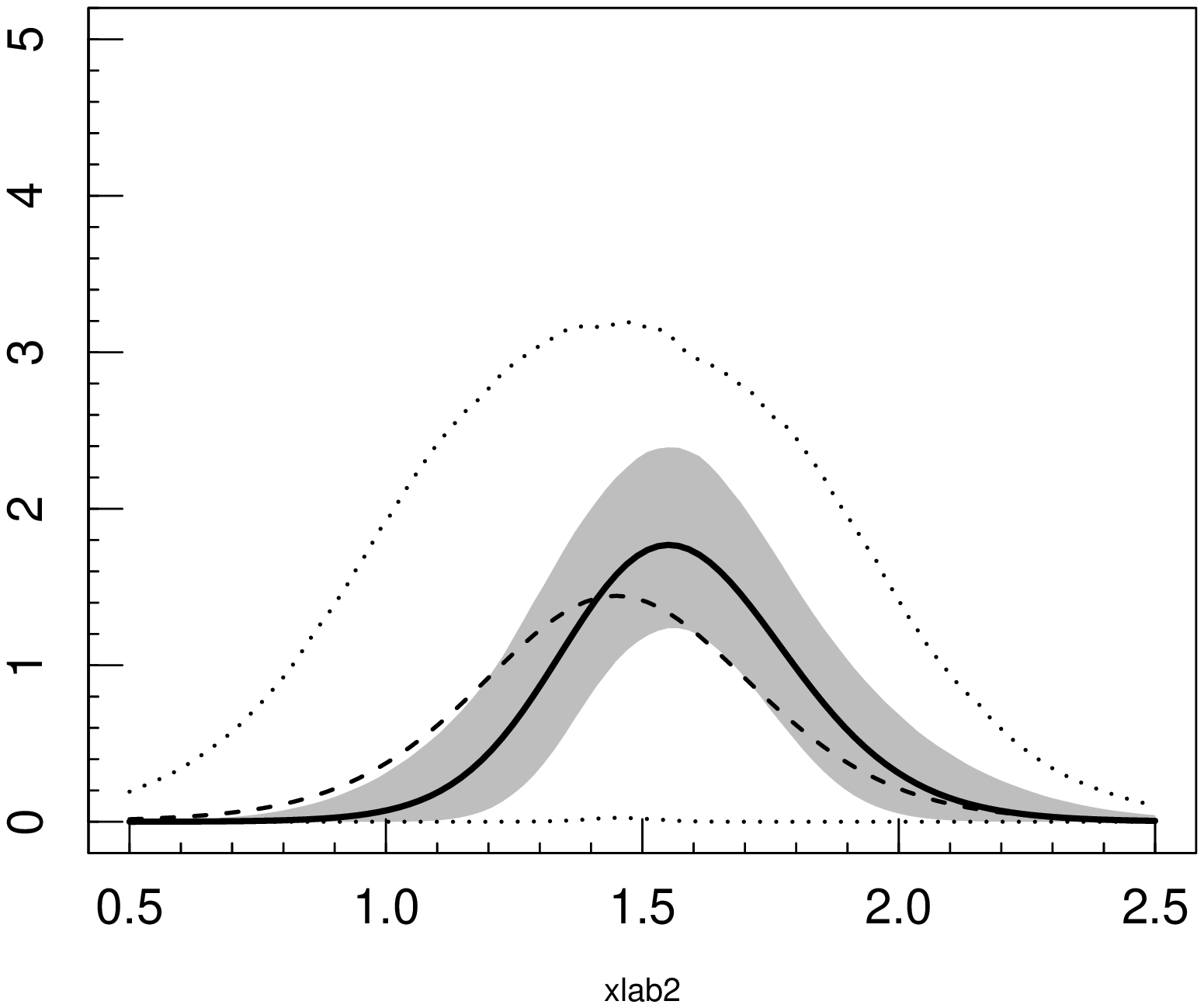} 	
\caption{Comparison of prior mean and 95\% interval estimates (dashed and
dotted lines) with posterior mean and 95\% interval estimates (solid lines
and gray bands) for the NS mass density corresponding to the DNS 
systems (left panel) and the NS-WD systems (right panel).
Refer to \Sref{inference} for further details.}
\label{Fig:densities-with-intervals}   
\end{figure}

It is noteworthy from Figure~\ref{Fig:densities} that the NS-WD systems 
posterior predictive density suggests positive skewness in the 
NS mass distribution. This is also reflected in the posterior
distribution for the skewness parameter $\alpha$ given the NS-WD 
systems data; specifically, the posterior mean for $\alpha$ is $0.90$
and $\text{Pr}(\alpha > 0 \mid \text{data}) = 0.78$. 
In contrast, the corresponding results for the DNS systems data
are $\text{E}(\alpha \mid \text{data}) = -0.03$
and $\text{Pr}(\alpha > 0 \mid \text{data}) = 0.49$ supporting 
symmetry (and normality) for the DNS systems mass distribution.

Next, we supplement the point estimates in Figure~\ref{Fig:densities} 
with uncertainty bands for the NS mass density. To this end, using 
a grid of mass values in $0.5 \, \mathrm{M}_{\odot}$ to 
$2.5 \, \mathrm{M}_{\odot}$, we evaluate the skewed normal NS mass
density in (\ref{skew-normal}) at each of the posterior samples for 
its parameters $(\mu,\sigma,\alpha)$. This produces a sample of 
densities which can be averaged to obtain the posterior mean 
NS mass density estimate, given by the solid lines in 
Figure~\ref{Fig:densities-with-intervals}. (Formally, the posterior
mean estimate is equivalent to the posterior predictive density, and 
thus the solid lines in Figure~\ref{Fig:densities-with-intervals}
agree with the estimates in Figure~\ref{Fig:densities}.)
However, we can now also depict the posterior uncertainty for the 
entire NS mass density through percentiles from the posterior sample 
of densities. In Figure~\ref{Fig:densities-with-intervals}, we use
the 0.025 and 0.975 percentiles, and thus the gray bands depicting 
the posterior uncertainty correspond to 95\% interval estimates
for the NS mass density.

Finally, Figure~\ref{Fig:densities-with-intervals} plots also the prior 
point and 95\% interval estimates for the NS mass density. These
are produced as discussed above for the posterior inference results, 
but in this case using samples from the prior distribution assigned to 
the model parameters. Hence, Figure~\ref{Fig:densities-with-intervals} 
shows the prior guess at the NS mass density (the prior mean estimate
given by the dashed line) as well as the extent of variability in the
prior (encapsulated by the dotted lines). This provides an effective
means to summarize the extent of prior information for the NS mass
density incorporated into the model through the specific priors
for the model parameters. Moreover, the comparison with the 
corresponding estimates given the data illustrates the amount of 
{\it prior-to-posterior} learning, which is evidently significant for 
both the DNS and NS-WD systems.

\subsection{Model Checking}
\label{Sec:checking}

The predictive performance of the statistical model developed in 
Sections \ref{Sec:model} and \ref{Sec:inference} was evaluated using 
a well-founded technique for Bayesian model checking
(see, for example, Chapter 6 of \citealt{Gelman:03}).
Briefly, for each data point $(m_{i},c_{i},d_{i})$, we obtained the
posterior predictive distribution,
$\mathcal{P}(m_{i}^{rep} \mid \text{data})$, for {\it replicated response} 
$m_{i}^{rep}$, that is, the pulsar mass estimate that we would observe 
if the experiment that produced the data was to be replicated. Details 
on sampling from these posterior predictive distributions are included 
in Appendix \Sref{postpred}. An indication of how well the model is performing 
predictively can be obtained by checking where the observed pulsar
mass estimate $m_{i}$ lies within the corresponding posterior
predictive density. A relatively large number of observations falling 
in the tails of the respective predictive densities is indicative of lack 
of model fit. Under the proposed model, all 
pulsar mass estimates from both the DNS and NS-WD systems were 
effectively captured within their corresponding posterior predictive
distributions; see the plots in Appendix \Sref{postpred}. These results provide a further 
illustration of the predictive power of the model as it {\it replicates}
the appropriate type of asymmetry for the responses with asymmetric 
measurement errors.

\section{Summary}\label{Sec:summ}

We overview the physical processes that tune masses of NSs in \Sref{theo}. In order to theoretically estimate the viable range for NS masses, we derive the birth mass (\Sref{birth}, $M_{birth}=1.08$--$1.57\msun $) and the amount of mass expected to be transferred onto recycled NSs during the binary phase (\Sref{accr}, $\Delta m_{acc}\approx 0.1$--$0.2\msun$). We then discuss why the constraints on the maximum NS mass ($M_{max}=1.5$--$3.2\msun $) are less stringent and comment on the sources of uncertainties in \Sref{maxi}.

In order to maintain a uniform approach in our analysis, we refrained from including additional constraints that may arise from assumptions such as the possible relationship between the binary period and the mass of the remnant white dwarf (i.e., $P_{b}-m_{2}$ relationship) suggested by \cite{Rappaport:95}. While more elaborate and hierarchical implementation methods may be utilized in deducing implication of other assumptions, a use of more inclusive approaches may only convolute the mass inference, which is contrary to the goal of this work. Throughout our analysis, we only assume that Einstein's prescription for general relativity is correct and include mass measurements which are considered secure (\Sref{obs}).

We then subject the pulsar mass measurements to a detailed statistical analysis. In \Sref{esti} we show that a Bayesian approach offers an effective means for inference (for detailed derivation, see Appendix \Sref{inf}), using a model that accommodates skewness in distributions for both the underlying NS masses and the measurement errors.

\section{Discussion and Conclusions}\label{Sec:disc}

\subsection{Previous Studies}\label{Sec:prev}

The first article that reviewed pulsar mass measurements in order to deduce the range of masses NSs can attain, was published by \cite{Joss:76}. They used mass measurements from 5 sources (PSR B1913+16, Her X-1, Cen X-3, SMC X-1, and 3U0900$-$40), which were predominantly X-ray sources, and found a marginally consistent range of $1.4$--$1.8\msun$. 

\cite{Finn:94} used a Bayesian statistical approach for the first time to infer limits on the NS mass distribution. By using the mass measurements of 4 radio pulsars (PSRs B1913+16, B1534+12, B2127+11C, and B2303+46), he concluded that NS masses should fall mainly in the range between $1.3$--$1.6\msun$. 

A comprehensive paper on pulsar masses was published by \cite{Thorsett:99}. Their analysis based on 26 sources yielded a remarkably tight mass range at $1.38^{+0.06}_{-0.10}\msun$. The width of their NS mass inference was mainly driven by the narrow error bands of the DNS mass measurements. 

The recent work by \cite{Schwab:10} analyzes masses of 14 sources with an approach based on comparing the cumulative distribution function (CDF) with an idealized Gaussian. It is well understood that the K-S test should be used with caution in cases where deviations occur in the tails \citep{Mason:83}. Additionally, even in data samples where the number of outliers in the tails are considerably larger and associated measurement errors are taken into account, a K-S approach will still remain limited in quantifying the significance of the outliers. Therefore, while the bimodal feature found for the initial mass function (i.e. $M_{birth}$) may be consistent with theoretical expectations for remnant masses produced by electron-capture versus Fe-core collapse SNe \citep{Podsiadlowski:04}, the evidence for a deviation of $M_{birth}$ from a unimodal distribution is still tentative. In order to firmly establish a potential multi-modal feature for the NS birth mass distribution, a more diverse sample tested with statistics that allows sensitivity and performance checks are required. 

\subsection{Maximum Mass Limit}\label{Sec:maxx}

Our model is flexible and sensitive enough for detecting signatures of a potential truncation in the underlying mass distribution. We pay particular attention to whether there are signatures of truncation in the implied mass distribution at the high mass end for two reasons: 1) A high mass limit set by the EOS of the NS matter, or by general relativity, should produce a relatively sharp cut off, which will manifest itself as a truncation. 2) The nature of potential skewness of the underlying mass distribution will provide insight into how the NSs are produced and the evolutionary link between the two NS populations. 

The masses of NSs in DNS systems imply a tight symmetric distribution while the distribution of masses of NSs in NS-WD systems show evidence of skewness with a heavy tail on the high mass end. On the other hand, both populations show no evidence of a strong truncation limit on either end.

These have important implications: The stochastic nature of evolutionary processes such as long term stable accretion naturally produces a wider distribution for NSs in NS-WD systems. This along with the lack of a strong truncation indicate that, in particular, the high mass end of the NS mass distribution is driven by evolutionary constraints. As a result, this rules out the possibility that an upper mass limit is set by the EOS of matter or general relativity for NSs with masses M $\lesssim$ 2.1$\msun$. Therefore, the 2.1$\msun$ upper mass limit implied by NSs in NS-WD systems should be considered as a minimum secure limit to the maximum NS mass rather than an absolute upper limit to NS masses. The heavy tail of the mass distribution of NSs in NS-WD systems favors the possibility that at least some of these pulsars are born as massive NSs.

\subsection{Central Density and the Equation of State}

All EOSs that require a maximum NS mass $M_{max}\le2.1\msun$ are ruled out. The implied stiffness of the EOS largely precludes the presence of meson condensates and hyperons at supranuclear densities. Consequently, lower central densities, larger radii and thicker crusts for NSs are favored \citep{Shapiro:83}.

The energy density-radius relation implied by \cite{Tolman:39}, when combined with the causality limit, gives an analytical solution for an upper limit on the central density
\bee
\rho_{c}M^{2}=15.3 \times 10^{15} \msun^{2} \,\text{g}\,\text{cm}^{-3}.
\eee 
With a 2.1$\msun$ secure lower limit on the maximum NS mass, we set a 95\% confidence upper limit to the central density of NSs, which is 
\bee
\rho_{max}<3.47\times10^{15} \text{g}\,\text{cm}^{-3}
\eee 
corresponding to $\approx$11$\rho_{s}$ for a fiducial saturation threshold $n_{s}\sim 0.16$ fm$^{-3}$.

Exotic matter such as hyperons and Bose condensates significantly reduce the maximum mass of NSs. Therefore, a strict lower limit on the maximum NS mass M$_{max}>2.1\msun$ rules out soft EOSs with extreme low density softening 
and require the existence of exotic hadronic matter \citep[see][for review]{Lattimer:07}. NSs with deconfined strange quark matter mostly have maximum predicted masses lower than 2.1$\msun$. Hence, EOSs with strange quark matter that predict maximum masses smaller than 2.1$\msun$ can also be ruled out as viable configurations for NS matter.

\subsection{Evidence for Alternative Evolution and the Formation of Massive Neutron Stars?}

A 2.1$\msun$ upper limit to masses of NSs in NS-WD system poses a problem. If all millisecond pulsars were indeed NSs that are recycled from a first generation of normal pulsars, the implied distribution should be consistent with a recycled version of the initial mass distribution. While the peaks of the distributions for DNS and NS-WD systems appear to be consistent with the expectations of standard recycling (\Sref{accr}), the widths imply otherwise. As shown in \Fref{psr}, $\Delta m_{acc}=0.22\msun$ lies within the expected range. However, with typical accretion rates experienced during the LMXB phase ($\dot{m}_{acc} \sim 10^{-3}\,\dot{\text{M}}_{\text{Edd}}$), it is difficult to accumulate sufficient mass onto NSs that started their lives as a $\sim$1.4$\msun$ NS and produce NSs with masses $\sim$2.1$\msun$ such as PSR B1516+02B. Even with initial masses of $\sim$1.6$\msun$ these sources need to accrete $\Delta m \approx0.4-0.5\msun$ during their active accretion phase. This requires long term stable active accretion at unusually high rates. 

Based on the \ppd demographics of millisecond pulsars, \cite{Kiziltan:09} argue that $\approx$ 30\% of the millisecond pulsar population may be produced via a non-standard evolutionary channel. This prediction falls in line with a distribution that has a consistent recycled peak but has an unusual width (and some skewness with a heavy tail on the high mass end) which extends up to 2.1$\msun$. While it is difficult to quantify the formation rate(s) of non-standard processes that may produce these NSs, it is clear that the standard scenario requires at least a revision. Such a revision should consistently reconcile for the observed \ppd distribution of millisecond pulsars, along with the long term sustainability of unusually high accretion rates that is required to produce the second generation of massive NSs. 

The only viable alternative to a major revision of the mass evolution implied by the standard recycling scenario, also corroborated by the lack of truncation of the underlying NS mass distribution and enforced by the heavier tail of the skewed distribution inferred from NS-WD systems, is then to form massive NSs.

\acknowledgements 
The authors thank P. Freire for sharing updated pdf's from which some of the NS mass estimates were extracted in \Tref{nswd}. B.K. and S.E.T. acknowledge NSF grant AST-0506453. The authors thank the anonymous referee for a critical review. After this work was submitted for initial review, \cite{Ozel:12} discussed an alternate approach for estimating the NS mass distribution assuming a Gaussian underlying distribution.

\bibliographystyle{apj}	

\appendix
%
%
%
%
\phantomsection \label{Sec:appendix}

\section{Bayesian Inference for the Neutron Star Mass Distribution}\label{Sec:inf}

\subsection{Hierarchical Model Formulation}
\label{model}

The skewed normal distribution in (\ref{skew-normal}) admits a stochastic 
representation as a mixture of normal distributions with a truncated normal 
mixing distribution. Introducing truncated normal latent random variables $z_{i}$ and 
reparameterizing $\text{SN}(\cdot \mid \mu, \sigma, \alpha)$ to
$\text{SN}(\cdot \mid \mu, \tau^{2}, \psi)$ as described in \Sref{inference}, we have 
\begin{equation}\label{eqn:stochastic}
\mathcal{M}_{i}\mid z_i,\mu,\tau^2,\psi \stackrel{ind}{\sim}\text{N}(\mathcal{M}_{i} \mid \mu+\psi z_i,\tau^2);
\,\,\,\,\,\,\,
z_i \stackrel{ind}{\sim} \text{N}(0,1)1_{[0,\infty)}(z_i)
\end{equation}
as a hierarchical mixture representation of $\text{SN}(\mathcal{M}_{i} \mid \mu, \tau^{2}, \psi)$
(that is, if we marginalize over
the $z_{i}$ in (\ref{eqn:stochastic}), we obtain $\mathcal{M}_{i} \mid \mu,\tau^2,\psi \stackrel{ind}{\sim}$
$\text{SN}(\mathcal{M}_{i} \mid \mu, \tau^{2}, \psi)$). Here, 
$\text{N}(0,1)1_{[0,\infty)}(z_i)$ indicates a standard normal distribution restricted to $\mathbb{R}^+$.

To facilitate MCMC posterior simulation, we work with this formulation for the skewed 
normal distribution. The second line of the hierarchical model for the data given in 
(\ref{hierarchical-model}) is therefore changed as indicated in (\ref{eqn:stochastic}), 
to give the following hierarchical model for the data:
\[m_i \mid \mathcal{M}_i\stackrel{ind}{\sim} \text{AN}(m_i - \mathcal{M}_i \mid c_i,d_i),\quad i=1,\dots,n\]
\[\mathcal{M}_{i}\mid z_i,\mu,\tau^2,\psi \stackrel{ind}{\sim}\text{N}(\mathcal{M}_{i} \mid \mu+\psi z_i,\tau^2),\quad i=1,\dots,n\]
\[z_i\stackrel{ind}{\sim}\text{N}(z_i\mid 0,1)1_{[0,\infty)}(z_i),\quad i=1,\dots,n\]
The Bayesian model is completed with (conditionally conjugate) priors for the NS mass distribution 
parameters. Specifically, we place a bivariate normal prior, $\text{N}((\mu,\psi)^T\mid\theta,\Sigma)$,
on $(\mu,\psi)$ with mean vector $\theta$ and covariance matrix $\Sigma$, and an inverse-gamma 
prior, $\text{IG}(\tau^2|a,b)$, on $\tau^{2}$ with shape parameter 
$a>1$ and scale parameter $b$, such that the prior mean of $\tau^{2}$ is $b/(a-1)$.

\subsection{MCMC Posterior Simulation Method}
\label{post}

In addition to the NS mass distribution parameters $(\mu,\tau^2,\psi)$, the model parameters include the individual NS masses $\mathcal{M}_1,\dots,\mathcal{M}_n$, as well as the auxiliary random variables $z_1,\dots,z_n$. The likelihood function and the priors for 
$(\mu,\psi)$ and $\tau^2$ combine to give the posterior distribution for all model parameters which is proportional to: 
\begin{equation}\label{eqn:propto}
\text{N}((\mu,\psi)^T\mid\theta,\Sigma) \text{IG}(\tau^2|a,b)
\prod_{i=1}^{n} \{\text{AN}(m_i - \mathcal{M}_i \mid c_i,d_i)\text{N}(\mathcal{M}_i\mid \mu+\psi z_i,\tau^2)\text{N}(z_i\mid 0,1)1_{[0,\infty)}(z_i)\}.
\end{equation}
We utilize an MCMC posterior simulation algorithm to sample from the posterior distribution.
The MCMC algorithm dynamically updates the model parameters by simulating from their full 
conditional distributions in turn, and upon reaching convergence, the resulting samples are 
samples from the posterior distribution which may be used for inference. 
The key benefit of the augmented hierarchical model that includes the auxiliary variables 
$z_1,\dots,z_n$ is that it allows for the use of a Gibbs sampler, which is both efficient and 
straightforward to implement, since it involves sampling from standard distributions
as described below.

For each $i=1,...,n$, the full posterior conditional distribution for $\mathcal{M}_i$ is proportional to 
$\text{AN}(m_i - \mathcal{M}_i \mid c_i,d_i) \text{N}(\mathcal{M}_i\mid \mu+\psi z_i,\tau^2)$,
which can be shown to result in a mixture of two truncated normals. The first truncated normal has mean 
$\{(\mu+\psi z_i)c_i^2d_i^2+m_i\tau^2\}/(\tau^2+c_i^2d_i^2)$ and variance $(\tau^2c_i^2d_i^2)/(\tau^2+c_i^2d_i^2)$, and is restricted to the interval $(-\infty,m_i]$. The second truncated normal has mean $\{(\mu+\psi z_i)(d_i^2/c_i^2) + m_i\tau^2\}/(\tau^2+(d_i^2/c_i^2))$ and variance $(\tau^2d_i^2/c_i^2)/(\tau^2+(d_i^2/c_i^2))$, and is restricted to the interval $(m_i,\infty)$. 
The (unnormalized) weight of the first truncated normal is 
\[
\exp\left\{-\frac{(\mu+\psi z_i-m_i)^2}{2(\tau^2+c_i^2d_i^2)}\right\}\cdot\frac{\tau d_ic_i}{(\tau^2+d_i^2c_i^2)^{1/2}}\left\{\Phi\left(\frac{m_i-((\mu+\psi z_i)c_i^2d_i^2+m_i\tau^2)/(\tau^2+d_i^2c_i^2)}{\tau d_ic_i/(\tau^2+d_i^2c_i^2)^{1/2}}\right)\right\}\]
and that of the second is
\[
\exp\left\{-\frac{(\mu+\psi z_i-m_i)^2}{2(\tau^2+(d_i^2/c_i^2))}\right\}
\cdot\frac{\tau d_i/c_i}{(\tau^2+(d_i^2/c_i^2))^{1/2}}
\left\{1-\Phi\left(\frac{m_i-((\mu+\psi z_i)(d_i^2/c_i^2)+m_i\tau^2)/
(\tau^2+(d_i^2/c_i^2))}{(\tau d_i/c_i)/(\tau^2 + (d_i^2/c_i^2))^{1/2}}\right)\right\}.\]
Each $\mathcal{M}_i$ is therefore updated at each MCMC iteration by drawing from 
a mixture of two truncated normals, with parameters given above.

Next, the full conditional distribution for $z_i$ is proportional to 
$\text{N}(\mathcal{M}_{i} \mid \mu+\psi z_i,\tau^2)N(z_i\mid 0,1)1_{[0,\infty)}(z_i)$,
which results in a truncated normal distribution with mean 
$(\mathcal{M}_i-\mu)\psi/(\psi^2+\tau^2)$ and variance $\tau^2/(\psi^2+\tau^2)$, 
restricted to $[0,\infty)$.

The NS mass distribution parameters can be sampled with standard updates. 
The posterior full conditional distribution for $(\mu,\psi)$ can be derived as 
a bivariate normal distribution, 
\begin{equation}\label{eqn:fullcondmu} 
\text{N}\left( (\mu,\psi)^T\mid(\Sigma^{-1}+Z^TZ/\tau^2)^{-1}
(\Sigma^{-1}\theta+Z^T\boldsymbol{\mathcal{M}}/\tau^2),(\Sigma^{-1}+Z^TZ/\tau^2)^{-1}\right),
\end{equation}
where $\boldsymbol{\mathcal{M}}=(\mathcal{M}_1,\dots,\mathcal{M}_n)^T$, and $Z$ is 
an $n\times 2$ matrix with $Z_{i1}=1$ and $Z_{i2}=z_i$, for $i=1,\dots,n$. 
Lastly, the full conditional for $\tau^2$ is given by an inverse-gamma distribution with 
updated shape and scale parameters, 
$\text{IG}(a+n/2,b+\sum_{i=1}^{n}(\mathcal{M}_i-\mu-\psi z_i)^2/2)$.

The Gibbs sampler is therefore implemented with the following general procedure:
\begin{itemize}
\item Begin with initial values $(\{ \mathcal{M}_{1}^{(0)},...,\mathcal{M}_{n}^{(0)} \},
\{ z_{1}^{(0)},...,z_{n}^{(0)} \},\mu^{(0)},\psi^{(0)},\tau^{2(0)})$.
\item If the current sample at iteration $t$ is $(\{\mathcal{M}_{1}^{(t)},...,\mathcal{M}_{n}^{(t)} \},
\{z_{1}^{(t)},...,z_{n}^{(t)} \},\mu^{(t)},\psi^{(t)},\tau^{2(t)})$, obtain the next sample by simulating from the following distributions:
\begin{itemize}
\item[-]draw $\mathcal{M}_i^{(t+1)}$, $i=1,\dots,n$, from a mixture of truncated normal
distributions as described above, using the current values of parameters 
$\mu^{(t)}$, $\psi^{(t)}$, $\{ z_{1}^{(t)},...,z_{n}^{(t)} \}$ and $\tau^{2(t)}$
\item[-]draw $z_i^{(t+1)}$, $i=1,\dots,n,$ from a normal distribution with mean $(\mathcal{M}_i^{(t+1)}-\mu^{(t)})\psi^{(t)}/(\psi^{2(t)}+\tau^{2(t)})$ and variance $\tau^{2(t)}/(\psi^{2(t)}+\tau^{2(t)})$, restricted to the interval $[0,\infty)$
\item[-]draw $(\mu,\psi)^{(t+1)}$ jointly from the bivariate normal distribution
given in (\ref{eqn:fullcondmu}), using the current values of all other parameters
\item[-]draw $\tau^{2(t+1)}$ from $\text{IG}(a+n/2,b+\sum_{i=1}^{n}(\mathcal{M}_i^{(t+1)}-\mu^{(t+1)}-\psi^{(t+1)} z_i^{(t+1)})^2/2)$
\end{itemize}
\item Repeat the previous step, for $t=1,\dots,T$.
\end{itemize}

The chain appeared to converge very rapidly; when two chains were run with different 
initial values, they met within just a few iterations. There was some autocorrelation present 
for $\mu$ and $\psi$, but it fell below 0.2 after approximately a lag of 5, and samples 
were approximately uncorrelated by a lag of 10. In light of these results, all inferences are
based on 10,000 posterior samples, which came from a longer chain which was 
thinned, retaining every 10 posterior samples.

\subsection{Prior Specification}
\label{priors}

Hyperparameters $\theta=(\theta_\mu,\theta_\psi)^T$, $\Sigma$, $a$, and $b$ 
must be specified in this Bayesian model. First, note that $E(\mathcal{M})=$
$\theta_\mu+\theta_\psi E(z)$, and the skewness parameter of the NS mass
distribution is $\alpha=\psi/\tau$. To avoid favoring skewness in the prior 
distribution, we set $E(\psi)=\theta_\psi=0$, and then set $\theta_\mu$ to a 
reasonable prior location for the NS mass distribution for each system. The set 
of priors used for the results in \Sref{inference} had $\theta_\mu=1.3$ for the 
DNS system, and $\theta_\mu=1.45$ for the NS-WD system. With $\theta_\psi=0$, 
we have that $Var(\mathcal{M})=$ $b/(a-1)+\Sigma_{\mu}+\Sigma_{\psi}$, where 
$(\Sigma_{\mu},\Sigma_{\psi})$ are the elements of matrix $\Sigma$, which is taken
to be diagonal. We fix $a=3$ in each set of priors, and for the DNS system we set 
$b/(a-1)$, $\Sigma_{\mu}$, and $\Sigma_{\psi}$ all to $0.25^2/3$. For the NS-WD 
system, we set all three expressions to $0.3^2/3$. These result in a fairly dispersed
(thus, non-informative) prior distribution for NS masses as demonstrated by the 
prior interval estimates in \Fref{densities-with-intervals}.

\begin{figure}[t]
\centering
\includegraphics[height=6in, width=3.5in]{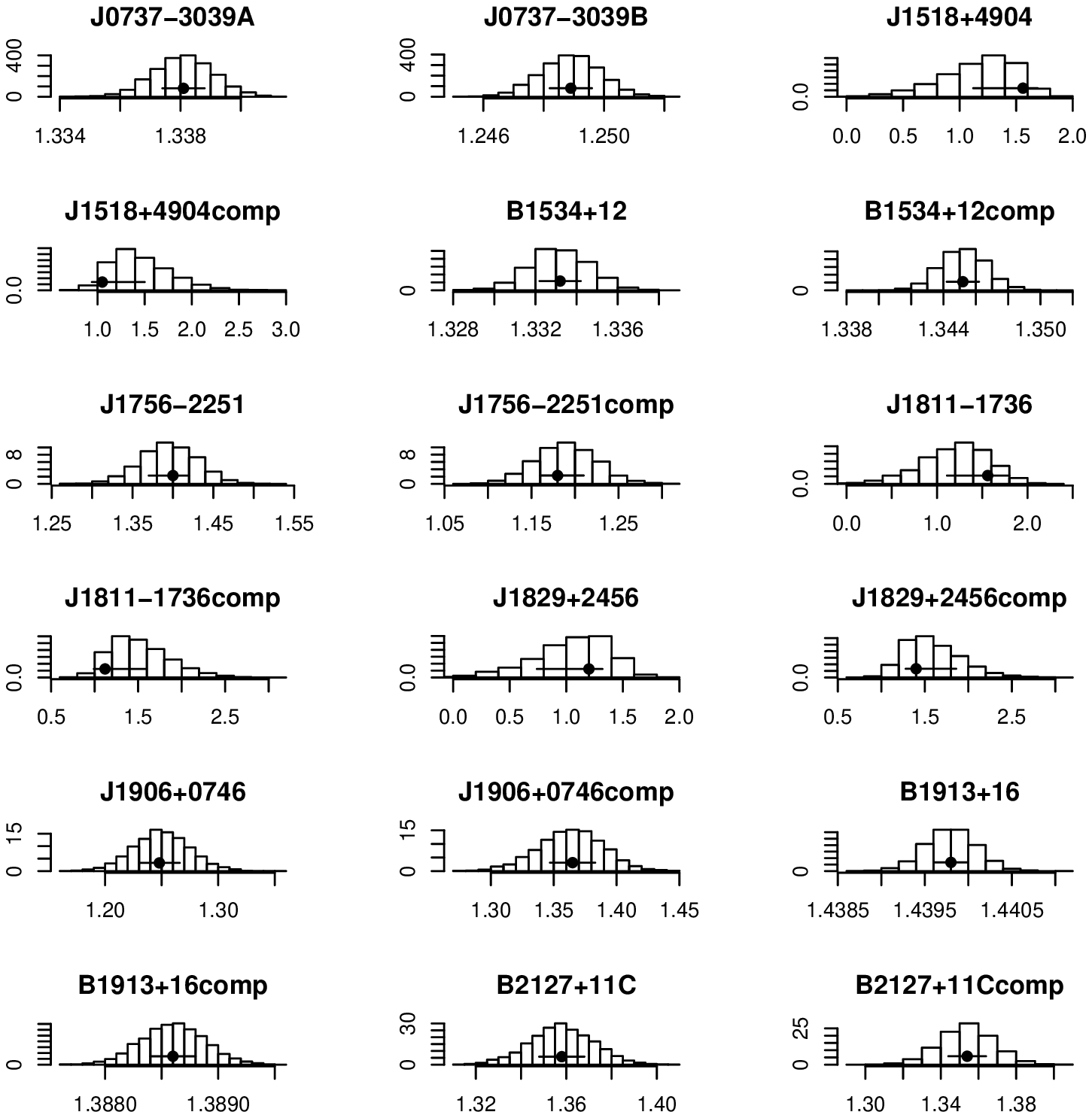}
\includegraphics[height=6in, width=3.5in]{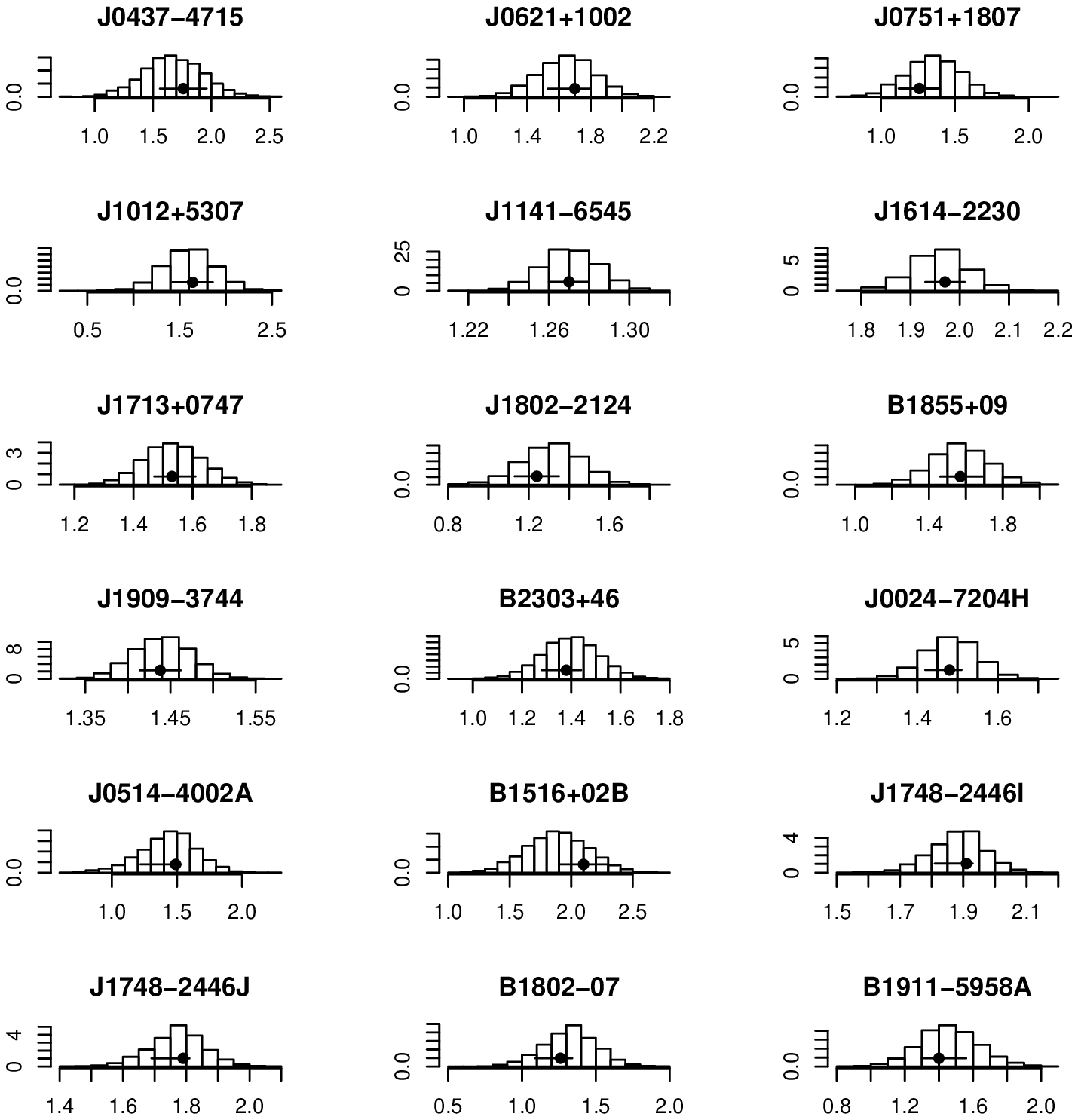}
\caption{Histograms of posterior predictive distributions 
for replicated responses for each pulsar. Columns 1-3 plot results for each observed pulsar mass 
estimate $m_i$ in DNS systems, and columns 4-6 for the NS-WD systems. The corresponding data for the measured NS masses, including both the estimates and error bars, are displayed in each panel.}
\label{Fig:appendix-fig}  
\end{figure}

\section{Posterior Predictive Model Checking Results}
\label{Sec:postpred}

As discussed briefly in \Sref{checking}, for model checking we work with 
the posterior predictive distributions for replicated responses for each pulsar. 
For each $i=1,...,n$, the corresponding expression is given by 
\[
\mathcal{P}(m_i^{rep}\mid \text{data}) = 
\int \text{AN}(m_i^{rep} - \mathcal{M}_i \mid c_i,d_i) p(\mathcal{M}_i \mid \text{data}) 
\, \text{d}\mathcal{M}_{i}.
\]
This distribution can be sampled by drawing a random variate from the 
asymmetric normal distribution in the integral representation above for 
each posterior sample of $\mathcal{M}_i$. \Fref{appendix-fig} provides histograms of these posterior predictive 
samples for replicate responses associated with each pulsar from the DNS 
and NS-WD systems. Again, the results suggest good predictive performance for the model.

\end{document}